\documentclass{elsart}
\usepackage{amsmath}
\usepackage{amssymb}
\usepackage{color,setspace,times}
\usepackage{amssymb,amsmath,graphicx,color,rotating,subfigure,url}
\usepackage{lineno}
\usepackage[square,sort&compress,comma]{natbib}

\journal{Physica A} 

\begin{document}

\begin{frontmatter}

\title{Multifractality in stock indexes: Fact or fiction? }
\author[BS,SS]{Zhi-Qiang Jiang},
\author[BS,SS,RCSE]{Wei-Xing Zhou\corauthref{cor}}
\corauth[cor]{Corresponding author. Address: 130 Meilong Road,
School of Business, P.O. Box 114, East China University of Science
and Technology, Shanghai 200237, China, Phone: +86 21 64253634, Fax:
+86 21 64253152.}
\ead{wxzhou@ecust.edu.cn} %


\address[BS]{School of Business, East China University of Science and Technology, Shanghai 200237, China}
\address[SS]{School of Science, East China University of Science and Technology, Shanghai 200237, China}
\address[RCSE]{Research Center of Systems Engineering, East China University of Science and Technology, Shanghai 200237, China}

\begin{abstract}
Multifractal analysis and extensive statistical tests are performed
upon intraday minutely data within individual trading days for four
stock market indexes (including HSI, SZSC, S\&P500, and NASDAQ) to
check whether the indexes (instead of the returns) possess
multifractality. We find that the mass exponent $\tau(q)$ is linear
and the singularity $\alpha(q)$ is close to 1 for all trading days
and all indexes. Furthermore, we find strong evidence showing that
the scaling behaviors of the original data sets cannot be
distinguished from those of the shuffled time series. Hence, the
so-called multifractality in the intraday stock market indexes is
merely an illusion.
\end{abstract}

\begin{keyword}
Econophysics, Multifractal analysis, Bootstrapping, Stock markets
\end{keyword}

\end{frontmatter}

\section{Introduction}


\label{s1:intro}

Econophysics is an emerging interdisciplinary field applying
concepts, theories, and tools borrowed from statistical physics,
nonlinear sciences, applied mathematics, and complexity sciences to
understand the complex self-organizing behaviors of financial
markets
\cite{Mantegna-Stanley-2000,Bouchaud-Potters-2000,Sornette-2003,Zhou-2007}.
This field has become to flourish since the pioneering work of
Mantegna and Stanley on the scaling behavior in the dynamics of the
Standard \& Poor's 500 index \cite{Mantegna-Stanley-1995-Nature},
which is closely related to the Pareto-L{\'{e}}vy law proposed by
Mandelbrot in the description of cotton price fluctuations
\cite{Mandelbrot-1963-JB}. Econophysicists have uncovered remarkable
similarities between financial markets and turbulent flows
\cite{Mantegna-Stanley-2000,Zhou-2007}. Such analogues include (but
not limited to) the evolution of probability densities of financial
returns \cite{Ghashghaie-Breymann-Peinke-Talkner-Dodge-1996-Nature}
based on the variational theory in turbulence
\cite{Castaing-Gagne-Hopfinger-1990-PD,Castaing-Gagne-Marchand-1993-PD,Castaing-Chabaud-Hebral-Naert-Peinke-1994-PB,Castaing-1994-PD},
inverse statistics in stock markets
\cite{Simonsen-Jensen-Johansen-2002-EPJB,Zhou-Yuan-2005-PA,Karpio-ZaluskaKotur-Orlowski-2007-PA}
motivated by the inverse structure function analysis of velocity
\cite{Jensen-1999-PRL,Biferale-Cencini-Vergni-Vulpiani-1999-PRE,Schmitt-2005-PLA,Pearson-vandeWater-2005-PRE,Zhou-Sornette-Yuan-2006-PD},
scale-invariant distribution of multipliers defined from volatility
of equities \cite{Jiang-Zhou-2007-PA} and from dissipating energy
\cite{Chhabra-Sreenivasan-1991-PRA,Chhabra-Sreenivasan-1992-PRL,Jouault-Lipa-Greiner-1999-PRE,Jouault-Greiner-Lipa-2000-PD},
and intermittency and multifractality of asset returns
\cite{Ghashghaie-Breymann-Peinke-Talkner-Dodge-1996-Nature,Mantegna-Stanley-1996-Nature}.

Indeed, the multifractal nature of equity returns is one of the most
important stylized facts. A small part of this literature contains
the studies on the foreign exchange rate
\cite{Ghashghaie-Breymann-Peinke-Talkner-Dodge-1996-Nature,Mantegna-Stanley-1996-Nature,Vandewalle-Ausloos-1998-IJMPC,Schmitt-Schertzer-Lovejoy-1999-ASMDA,Ivanova-Ausloos-1999-EPJB,Baviera-Pasquini-Serva-Vergni-Vulpiani-2001-PA,Muniandy-Lim-Murugan-2001-PA,Xu-Gencay-2003-PA},
gold price \cite{Ivanova-Ausloos-1999-EPJB}, commodity price
\cite{Matia-Ashkenazy-Stanley-2003-EPL}, returns of stock price or
indexes
\cite{Matia-Ashkenazy-Stanley-2003-EPL,Turiel-Perez-Vicente-2003-PA,Oswiecimka-Kwapien-Drozdz-Rak-2005-APP,Olsen-2000-PP,Turiel-Perez-Vicente-2005-PA,Norouzzadeh-Jafari-2005-PA,Bershadskii-2001-JPA,Andreadis-Serletis-2002-CSF,Gorski-Drozdz-Speth-2002-PA,Balcilar-2003-EMFT},
and so on. We note that the quantity {\em{price}} (or its logarithm)
in financial markets is the analogue of {\em{velocity}} in
turbulence. Similarly, the counterpart of {\em{velocity difference}}
in fluid mechanics is the asset {\em{return}}. In this framework, it
is natural that numerous multifractal analyses have been carried out
on the returns for financial equities similar to the velocity
differences for turbulent flows.

However, there are exceptions, where analysis is performed on
several indexes directly rather than their variations (the returns)
and the presence of multifractality in the several indexes is
claimed
\cite{Sun-Chen-Wu-Yuan-2001-PA,Sun-Chen-Yuan-Wu-2001-PA,Wei-Huang-2005-PA}.
Specifically, they performed multifractal analysis on the intraday
high-frequency data of Hang Seng Index (HSI), Shanghai Stock
Exchange Composite Index (SSEC), and Shenzhen Stock Exchange
Composite Index (SZEC) within individual trading days. The extracted
``multifractal'' spectra $f(\alpha)$ were then utilized to predict
abnormal price movements and serve as a risk measure in risk
management. It seems to us that a careful scrutiny on the obtained
multifractality should be undertaken based on the extremely narrow
spectra of the singularity $\alpha$. Two problems arise, casting
doubts on the aforementioned analysis \cite{Zhou-2007-JMSC}.

Firstly, based on the multifractal theory, there exists a constant
$\alpha(t)$ for each moment $t$ such that the investigated measure
$\mu$ on the neighbor $B(t,l)$ of $x$ scale with $l$ when the scale
$l \rightarrow 0$,
\begin{equation}
\mu\left(B(t,l)\right) \sim l^{\alpha(t)}.
 \label{Eq:PL}
\end{equation}
The measure $\mu$ is singular at arbitrary moment $t$ with the
singularity strength being $\alpha(t)$. When $\mu$ is defined as the
sum of index prices within a given time interval,
$\mu\left(B(t,l)\right)$ is approximately proportional to $l$, that
is, $\alpha(t) \approx 1$ for all $t$. This suggests that the
measure $\mu$ does not possess multifractal nature. This inference
is further supported by the fact that the span of singularity
strength $\Delta \alpha = \alpha_{\max} - \alpha_{\min} \approx 0$
in the real data
\cite{Sun-Chen-Wu-Yuan-2001-PA,Sun-Chen-Yuan-Wu-2001-PA,Wei-Huang-2005-PA}.

Secondly, in the analysis of multifractality in turbulence or
high-frequency financial data, the moment order $q$ should not be
greater than 8 in order to make the partition function converge.
Specifically, it is shown that the size of a time series should be
no less than one million to ensure the estimate of its eighth order
partition function statistically significant
\cite{Lvov-Podivilov-Pomyalove-Procaccia-Vandembroucq-1998-PRE,Zhou-Sornette-Yuan-2006-PD}.
The situation is similar for high-frequency financial data
\cite{Jiang-Zhou-2007-PA}. Hence, it is of little significance to
compute partition function for higher orders. In the analysis of
minutely (or five-minute) data within a time period of one day
\cite{Sun-Chen-Wu-Yuan-2001-PA,Sun-Chen-Yuan-Wu-2001-PA,Wei-Huang-2005-PA},
the size of the intraday high-frequency data is no more than 240
while the moment order is taken to be $-120 \leqslant q \leqslant
120$. This usually broad interval of $q$ casts further doubts on the
reported multifractality in the indexes.

Despite of the specific considerations discussed above, it is
worthwhile to put further comments in general on the investigation
of multifractality in financial data. The multifractal features in
financial series have attracted great interests, however, the origin
and significance of the extracted ``multifractality'' is less
concerned. On one hand, it has been shown that an exact monofractal
financial model can lead to an artificial multifractal behavior
\cite{Bouchaud-Potters-Meyer-2000-EPJB}. On the other hand, a time
series of the price fluctuations possessing multifractal nature
usually has either fat tails in the distribution or long-range
temporal correlation or both
\cite{Kantelhardt-Zschiegner-Bunde-Havlin-Bunde-Stanley-2002-PA}.
However, possessing long memory is not sufficient for the presence
of multifractality and one has to have a nonlinear process with
long-memory in order to have multifractality
\cite{Saichev-Sornette-2006-PRE}. In many cases, the null hypothesis
that the reported multifractal nature is stemmed from the large
fluctuations of prices cannot be rejected \cite{Lux-2004-IJMPC}.

In this work, we focus on the presence of multifractal feature in
stock market indexes and testing whether the obtained empirical
multifractality stems from random fluctuations. To address these
issues, we adopt the bootstrap approach by shuffling the intraday
index series and perform multifractal analysis on them. The results
are compared with that from original data. This paper is organized
as follows. In Sec.~\ref{s2:data}, we describe the data sets we
investigate. The basic multifractal method is explained in detail in
Sec.~\ref{s3:method}. Multifractal analysis of the data sets is
presented in Sec.~\ref{s4:MA}. Statistical bootstrapping tests are
conducted in Sec.~\ref{s5:STM}. Finally, Sec.~\ref{s6:con}
concludes.

\section{Data sets}
\label{s2:data}

To gain a more profound insight into the multifractality in intraday
stock market indexes, we investigate four important indexes,
{\it{i.e.}}, the Hang Seng Index (HSI), Shenzhen Stock Exchange
Composite Index (SZSC), Standard \& Poor's 500 Index (S\&P 500), and
the National Association of Securities Dealers Automated Quotation
(NASDAQ). HSI and SZSC are selected since they were used in the
original work of this topic
\cite{Sun-Chen-Wu-Yuan-2001-PA,Sun-Chen-Yuan-Wu-2001-PA,Wei-Huang-2005-PA}.
Both the Hongkong Stock Exchange and Shenzhen Stock Exchange are
emerging markets. The S\&P 500 and NASDAQ that are representative of
mature stock markets are chosen for comparison.

The data have been recorded at each minute in trading days. The HSI
index covers from Jan. 2, 1997 to May 28, 1997, the SZSC index is
from Nov. 12, 2001 to Aug. 17, 2006, the S\&P 500 index is recorded
from Jan. 2, 1997 to Feb. 26, 1999, and the NASDAQ index ranges from
Aug. 18, 2000 to Oct. 30, 2000. Eliminating the weekend, holidays
and, the days having recording errors, there are 101 days for the
HSI data, 1149 days for the SZSC data, 448 days for the S\&P 500
data, and 45 days for the NASDAQ data, respectively.

\section{Method}
\label{s3:method}

We use the box counting method following the work of
\cite{Sun-Chen-Wu-Yuan-2001-PA,Sun-Chen-Yuan-Wu-2001-PA,Wei-Huang-2005-PA}
to investigate the multifractal nature of the index series of each
trading day. Denote the intraday index series as $\{ I(t): t = 1, 2,
\cdot \cdot \cdot, T \}$, where $T = 240$ for HSI and SZSC, $T =
405$ for S\&P 500, and $T = 390$ for NASDAQ, respectively. For a
given box size $l$, we obtain $N = T/l$ boxes and construct a
measure $\mu$ on each box as follows,
\begin{equation}
\mu(n;l) = \mu\left([(n-1)l+1,nl]\right)= \sum_{i=1}^{l}
I[(n-1)l+i]~,
 \label{Eq:mu}
\end{equation}
where $[(n-1)l+1,nl]$ is the $n$-th box and $l \in [1, 2, 3, 4, 6,
10, 15, 20, 30, 40, 60, 80,\\ 120, 240]$ for HSI and SZSC, $l \in
[1, 3, 5, 9, 15, 27, 45, 81, 135, 405]$ for S\&P 500, and $l \in [1,
2, 3,5, 10, 13, 15, 26, 30, 39, 78, 130, 195, 390]$ for NASDAQ,
respectively. The sizes $l$ of the boxes are chosen such that the
number of boxes of each size is an integer to cover the whole time
series.

We then construct the partition function $\chi_q$ as
\begin{equation}
\chi_q(l) =  \sum_{n=1}^{N} \left[ \frac{\mu(n;l)}{\sum_{m=1}^{N}
\mu(m;l)} \right]^q~,
 \label{Eq:chi}
\end{equation}
and expect it to scale as
\begin{equation}
\chi_q(l) \sim l^{\tau(q)}~,
 \label{Eq:mu}
\end{equation}
which defines the exponent $\tau(q)$. The local singularity exponent
$\alpha$ of the measure $\mu$ and its spectrum $f(\alpha)$ are
related to $\tau(q)$ through a Legendre transformation
\cite{Halsey-Jensen-Kadanoff-Procaccia-Shraiman-1986-PRA}
\begin{equation} \label{Eq:alphaf}
\left\{ \begin{aligned}
         \alpha &= {\rm{d}}\tau(q)/{\rm{d}}q \\
                  f(\alpha)&=q \alpha -\tau(q)
                  \end{aligned} \right.~.
                          \end{equation}
In order to keep the comparability of our results with those in
\cite{Wei-Huang-2005-PA}, we also pose $-120 \leqslant q \leqslant
120$.

When $\mu(n;l)/\sum \mu(m;l) \ll 1$ and $q \gg 1$, the estimate of
the partition function $\chi$ will be very difficult since the value
is so small that it is out of the memory. To overcome this problem,
we can calculate the logarithm of the partition function $\ln
\chi_q(l)$ rather than the partition function itself. A simple
manipulation results in the following formula,
\begin{equation}
\ln \chi_q(l) = \ln \sum_{n=1}^{N} \left[
\frac{\mu(n;l)}{\max\limits_m\{\mu(m;l)\}}\right]^q + q \ln \left[
\frac{\max\limits_m\{\mu(m;l)\}}{\sum\mu(m;l)}\right]~,
 \label{Eq:logchi}
\end{equation}
where $\max\limits_m\{\mu(m;l)\}$ is the maximum of $\mu(m;l)$ for
$m=1, 2, \cdots, N$.

\section{Multifractal analysis}
\label{s4:MA}

\begin{figure}[htb]
\centering
\includegraphics[width=6cm]{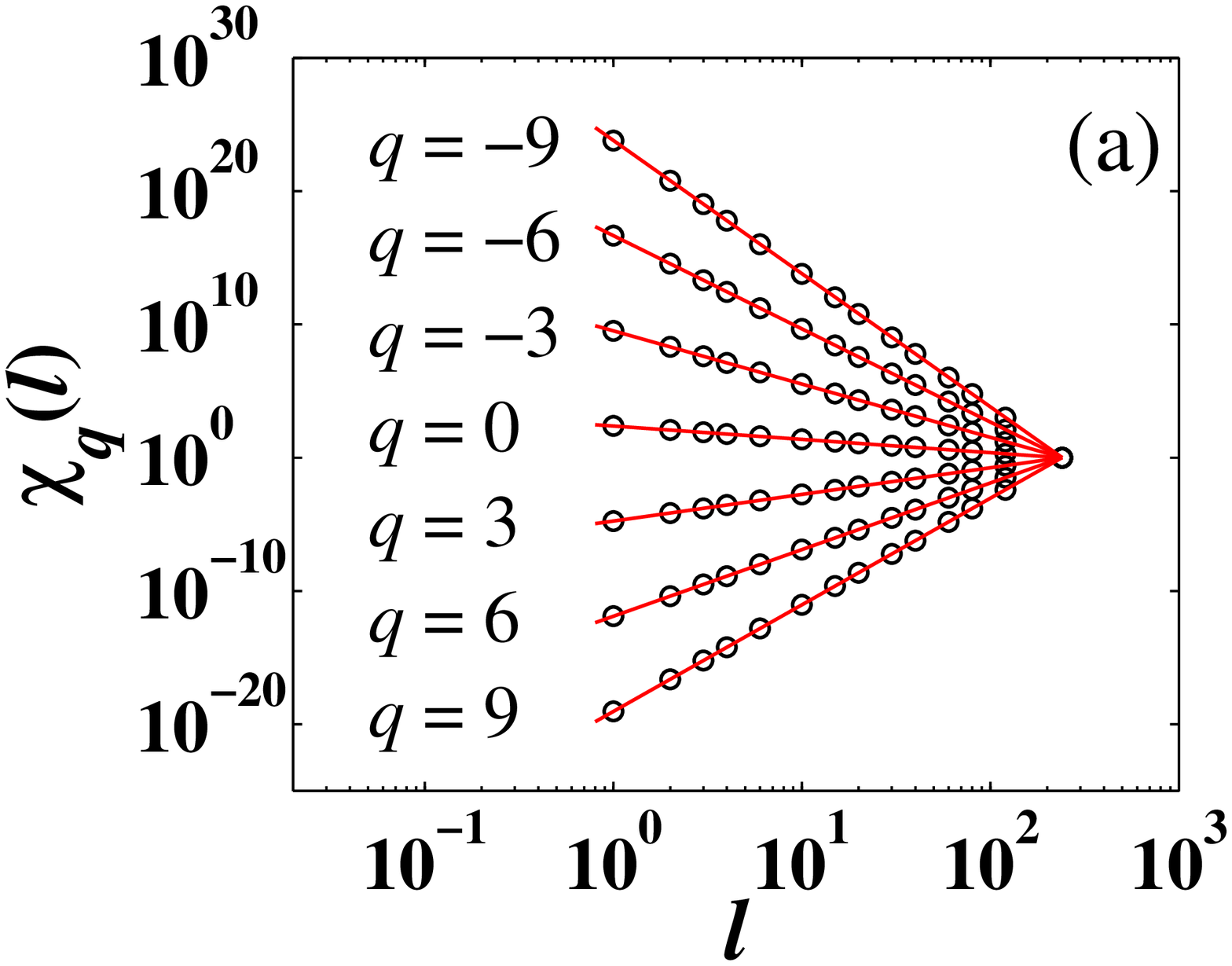}
\includegraphics[width=6cm]{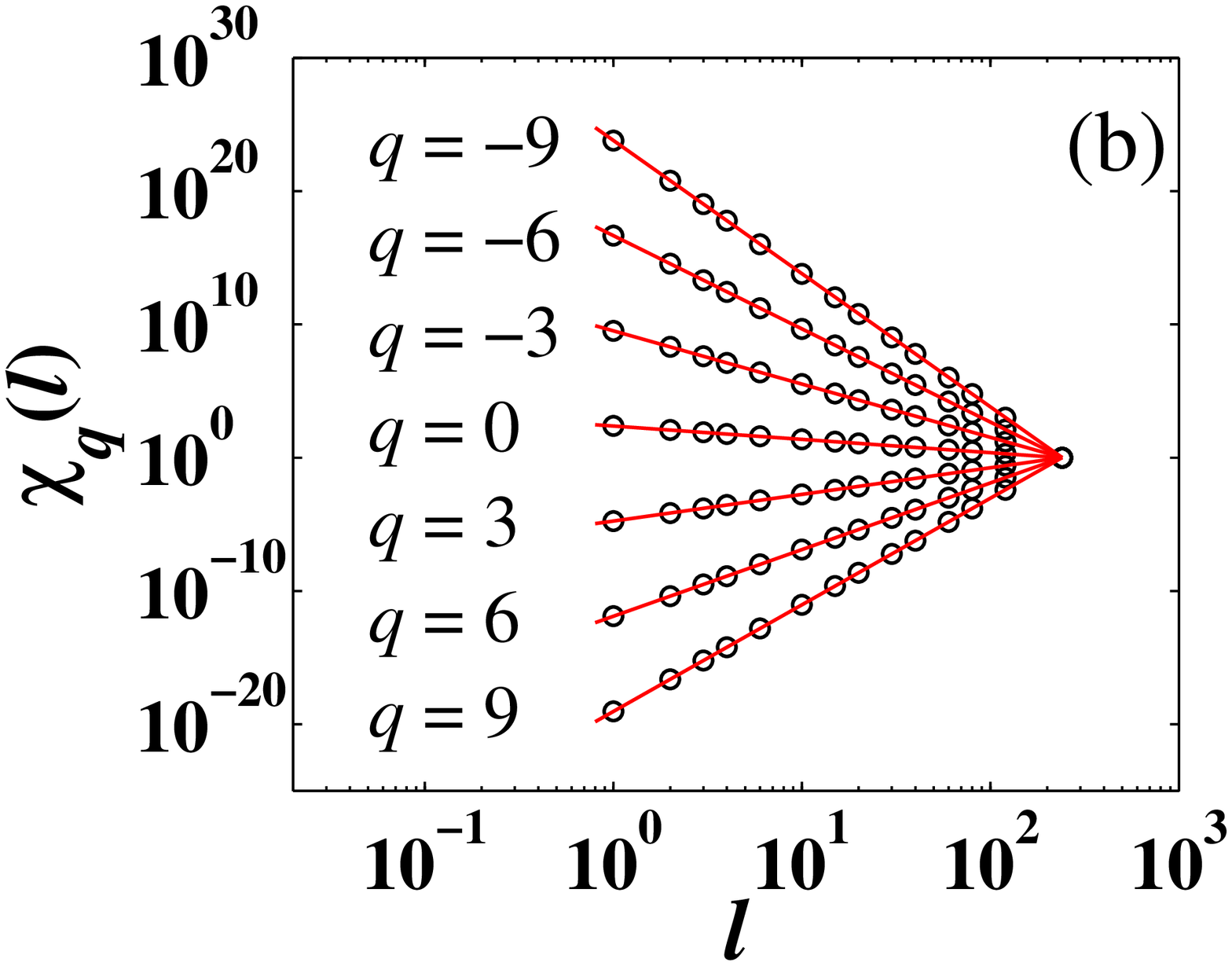}
\includegraphics[width=6cm]{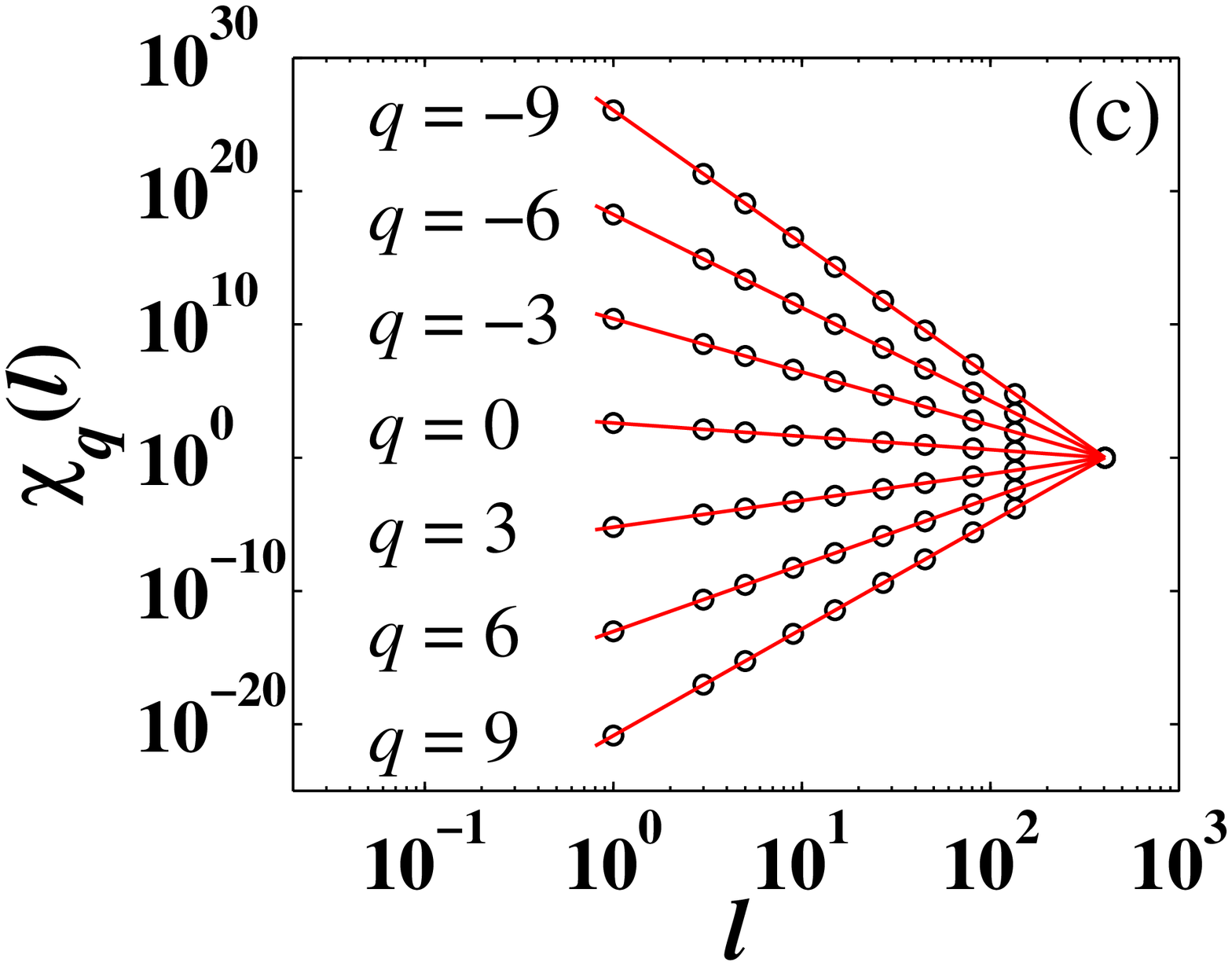}
\includegraphics[width=6cm]{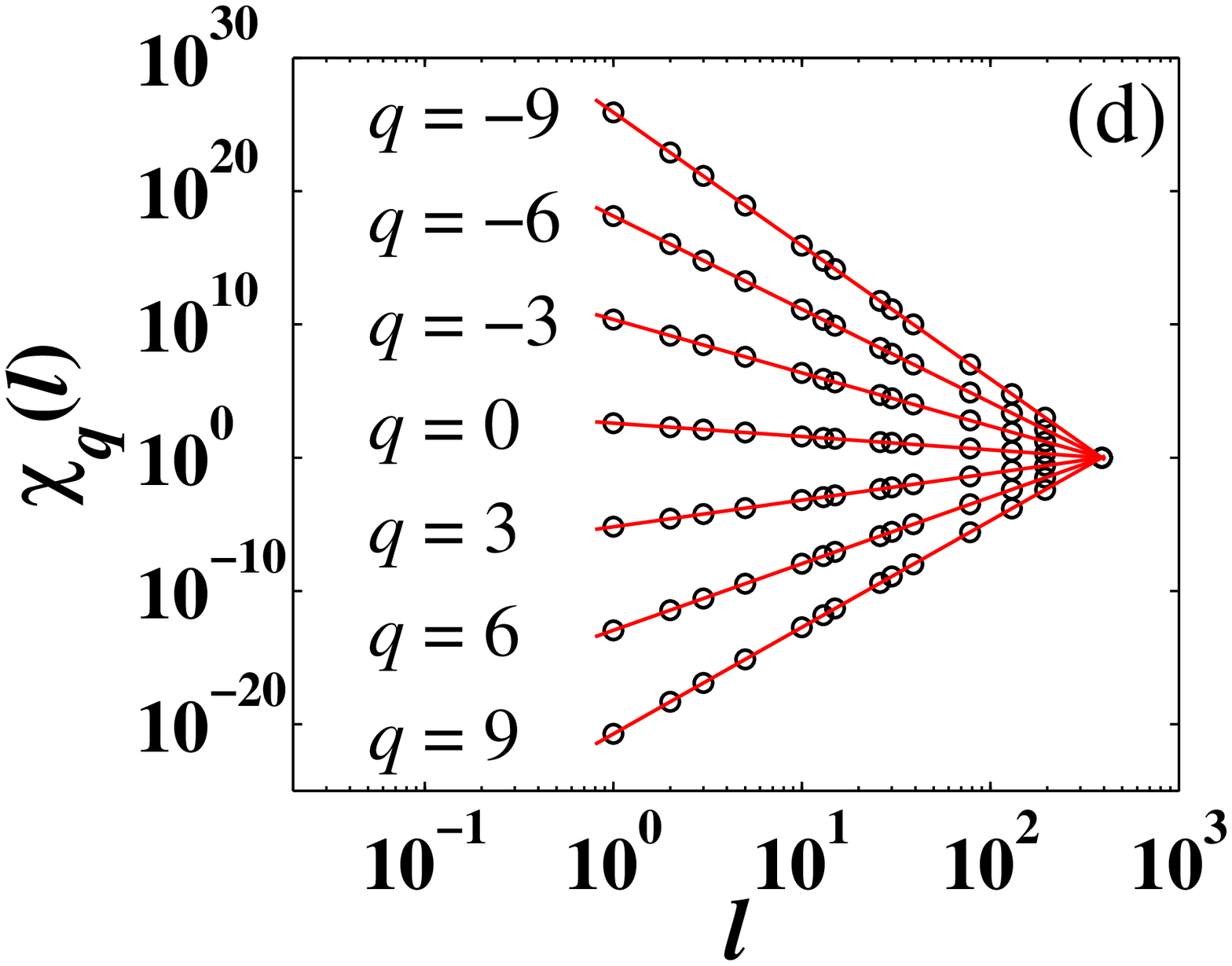}
\caption{Plots of $\chi_q(l)$ as a function of the box size $l$ for
different values of $q$ in log-log coordinates. The solid lines are
the least-squares fits to the data using linear regression (in
log-log coordinates) corresponding to power laws. (a) HSI, (b) SZSC,
(c) S\&P 500, and (d) NASDAQ.} \label{Fig:Chil}
\end{figure}

Four dates (Jan. 8, 1997 for HSI, Nov. 26, 2001 for SZSC, Feb. 10,
1997 for S\&P 500, and Aug. 22, 2000 for NASDAQ) are taken as
examples to show the results of multifractal analysis.
Figure~\ref{Fig:Chil} shows the dependence of the partition function
$\chi_q(l)$ on the box size $l$ for different values of $q$ in
log-log coordinates. Excellent power-law scaling of $\chi_q(l)$ with
respect to $l$ has been observed and the scaling range covers all
the selected values of $l$. The solid lines are the best linear fits
to the data.

\begin{figure}[htb]
\centering
\includegraphics[width=6cm]{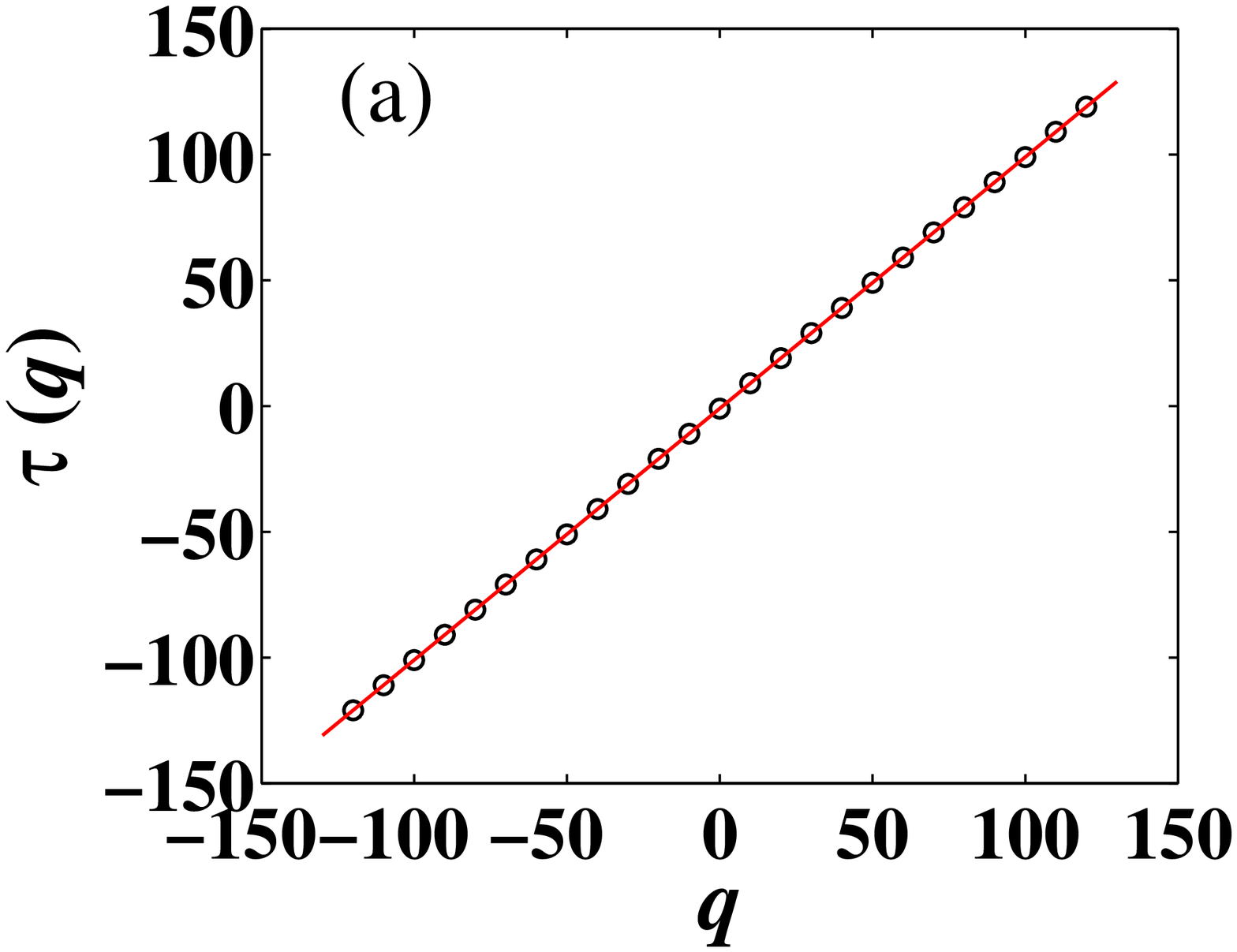}
\includegraphics[width=6cm]{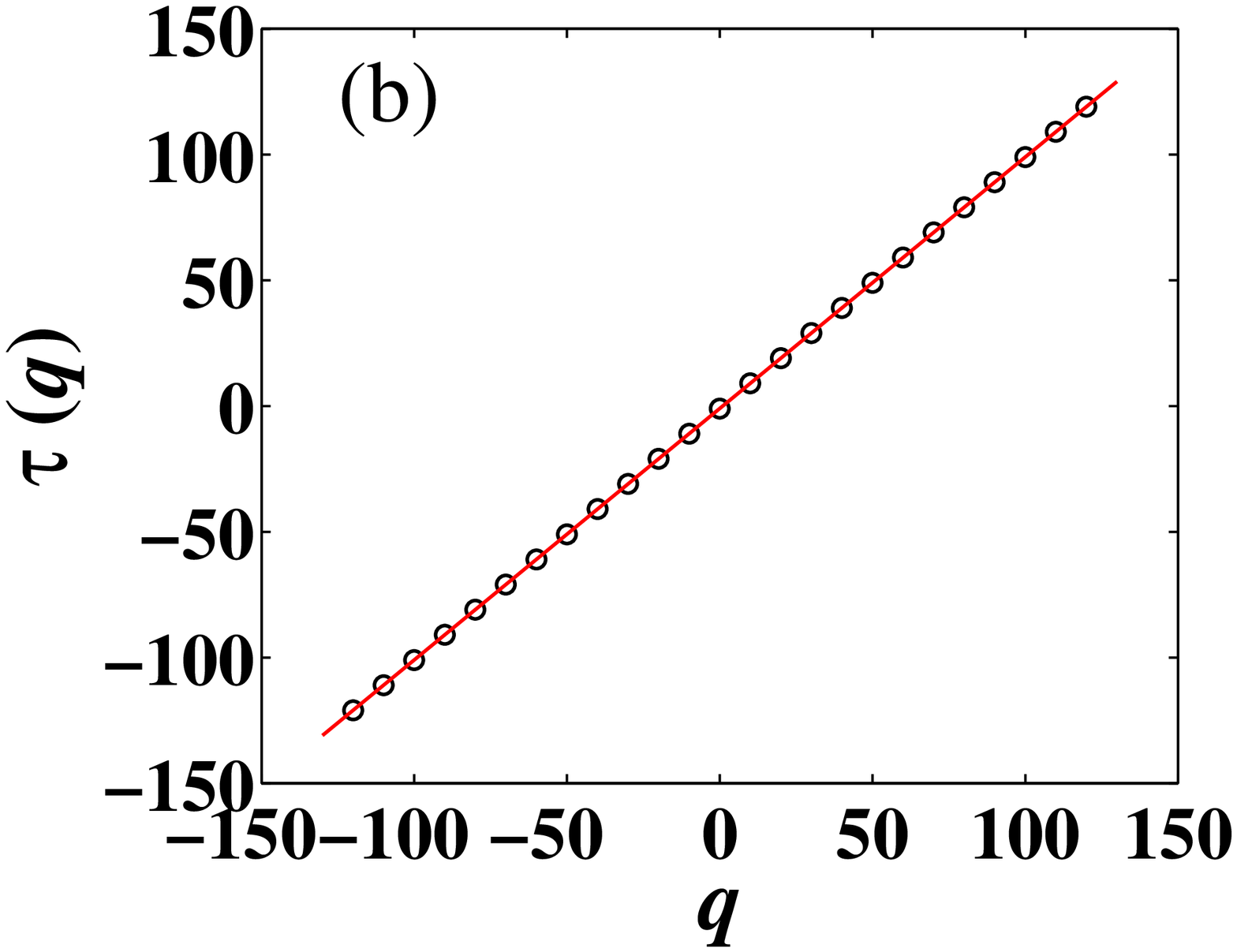}
\includegraphics[width=6cm]{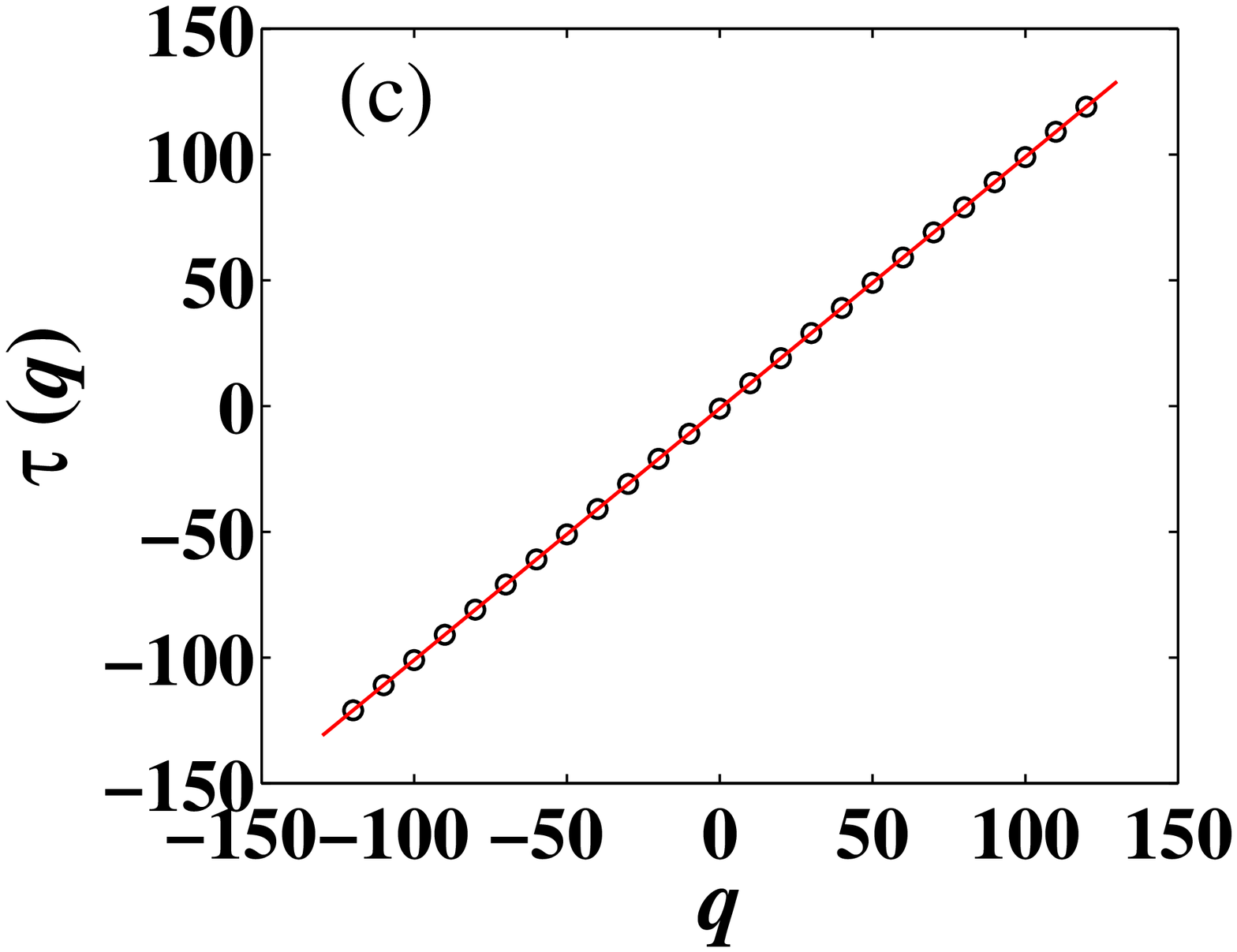}
\includegraphics[width=6cm]{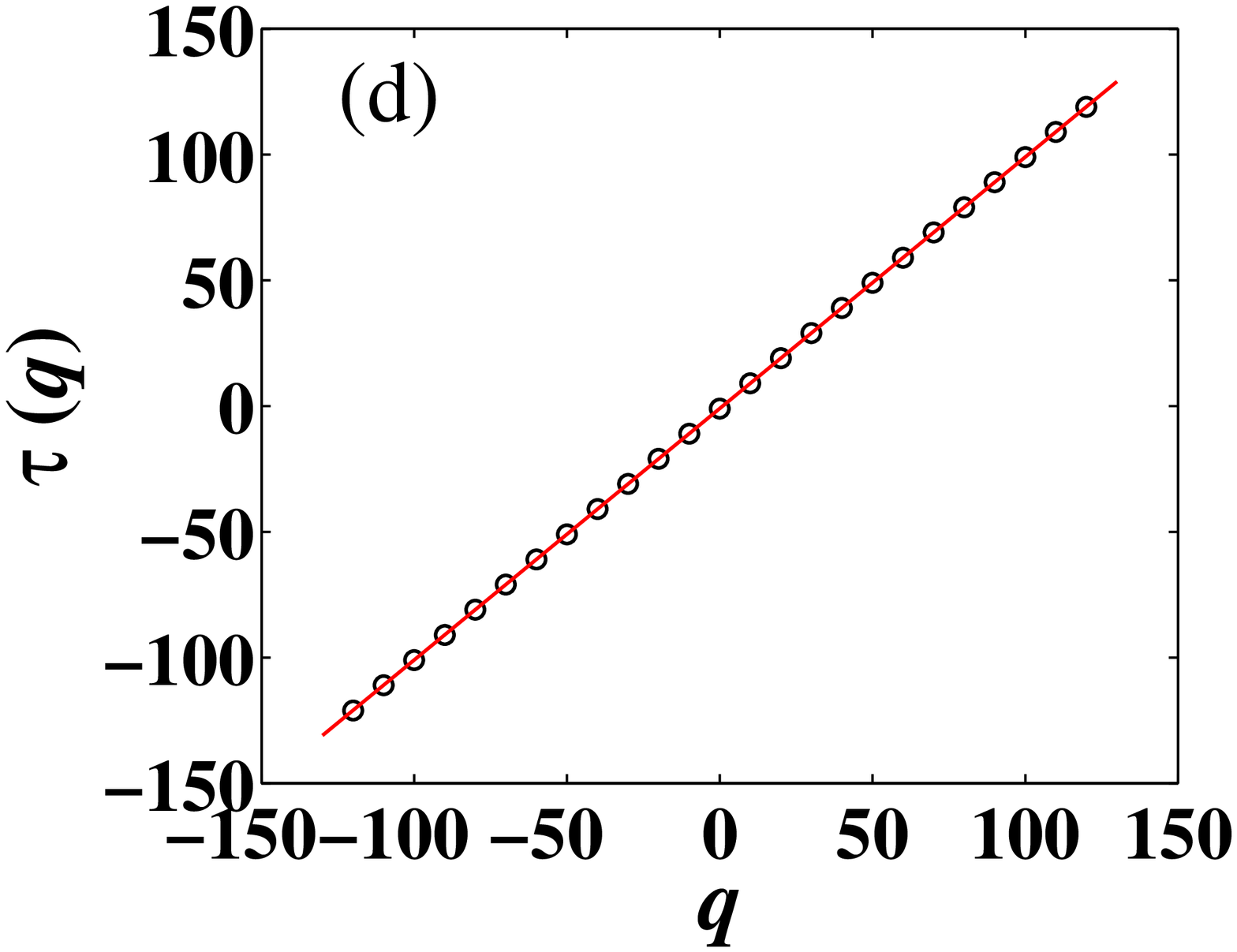}
\caption{Dependence of the scaling exponent $\tau(q)$ on the order
$q$. The solid lines are the least-squares fits to the data. (a)
HSI, (b) SZSC, (c) S\&P 500, and (d) NASDAQ.} \label{Fig:Tauq}
\end{figure}

The scaling exponents $\tau(q)$ are given by the slopes of the
linear fits to $\ln \chi_q(l)$ with respect to $\ln l$ for different
values of $q$. Figure~\ref{Fig:Tauq} plots the dependence of the
mass exponents $\tau(q)$ as a function of the moment order $q$. One
observes that there is an evident linear relationship between
$\tau(q)$ and $q$ for all the four examples. The solid lines are the
least-squares fits to the data. The slopes of the lines are
respectively $\bar\alpha = 1.000 \pm 0.001$ for HSI, $\bar\alpha =
1.000000 \pm 0.000003$ for SZSC, $\bar\alpha = 1.00000 \pm 0.00001$
for S\&P 500, and $\bar\alpha = 1.0001 \pm 0.0001$ for NASDAQ,
respectively. All the corresponding correlation coefficients of the
linear fits are equal to $1.0000$. Furthermore, the linear
relationships are also hold for other trading days. Therefore, there
is no evidence of nonlinearity in the functions $\tau$ and the
intraday stock market index do not exhibit multifractal nature.
Since $\alpha(q)={\rm{d}}\tau(q)/{\rm{d}}q$, we expect that
$\alpha(q)\approx1$ for all $q$, as expected in our discussion in
Sec.~\ref{s1:intro}.

\begin{figure}[htb]
\centering
\includegraphics[width=8cm]{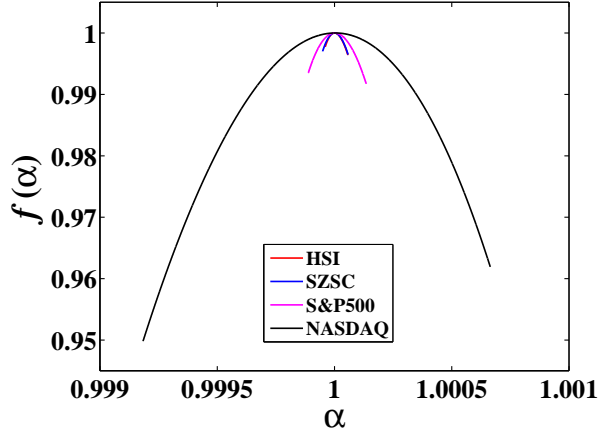}
\caption{(Color online) Multifractal spectra $f(\alpha)$ obtained by
the Legendre transform of $\tau(q)$ for different indexes.}
\label{Fig:Falpha}
\end{figure}

Figure~\ref{Fig:Falpha} presents the multifractal singularity
spectra $f(\alpha)$ obtained through Legendre transformation of
$\tau(q)$ defined by Eq.~(\ref{Eq:alphaf}). The curves in
Fig.~\ref{Fig:Falpha} have the geometrical features of the
conservable multifractal spectra \cite{Zhou-2001,Zhou-2007}, which
makes them look as if there is sound evidence for the presence of
multifractality. However, when looking at the disperseness of the
sigularity strength $\Delta \alpha\triangleq\alpha_{\max} -
\alpha_{\min}$, we find that $\Delta\alpha$ is  very close to zero.
It is well-known that $\Delta\alpha$ is an important parameter
qualifying the width of the extracted multifractal spectrum. The
larger is the $\Delta \alpha$, the stronger is the multifractality.
According to Fig.~\ref{Fig:Falpha}, even in the case of $-120
\leqslant q \leqslant 120$, $\Delta \alpha < 0.002$ for NASDAQ. One
can see that the values of $\Delta \alpha$ for other indexes are
much smaller than that of NASDAQ. This observation indicates that
there is no multifractality in stock market indexes.

\section{Statistical tests for multifractality}
\label{s5:STM}

\begin{figure}[htb]
\centering
\includegraphics[width=6cm]{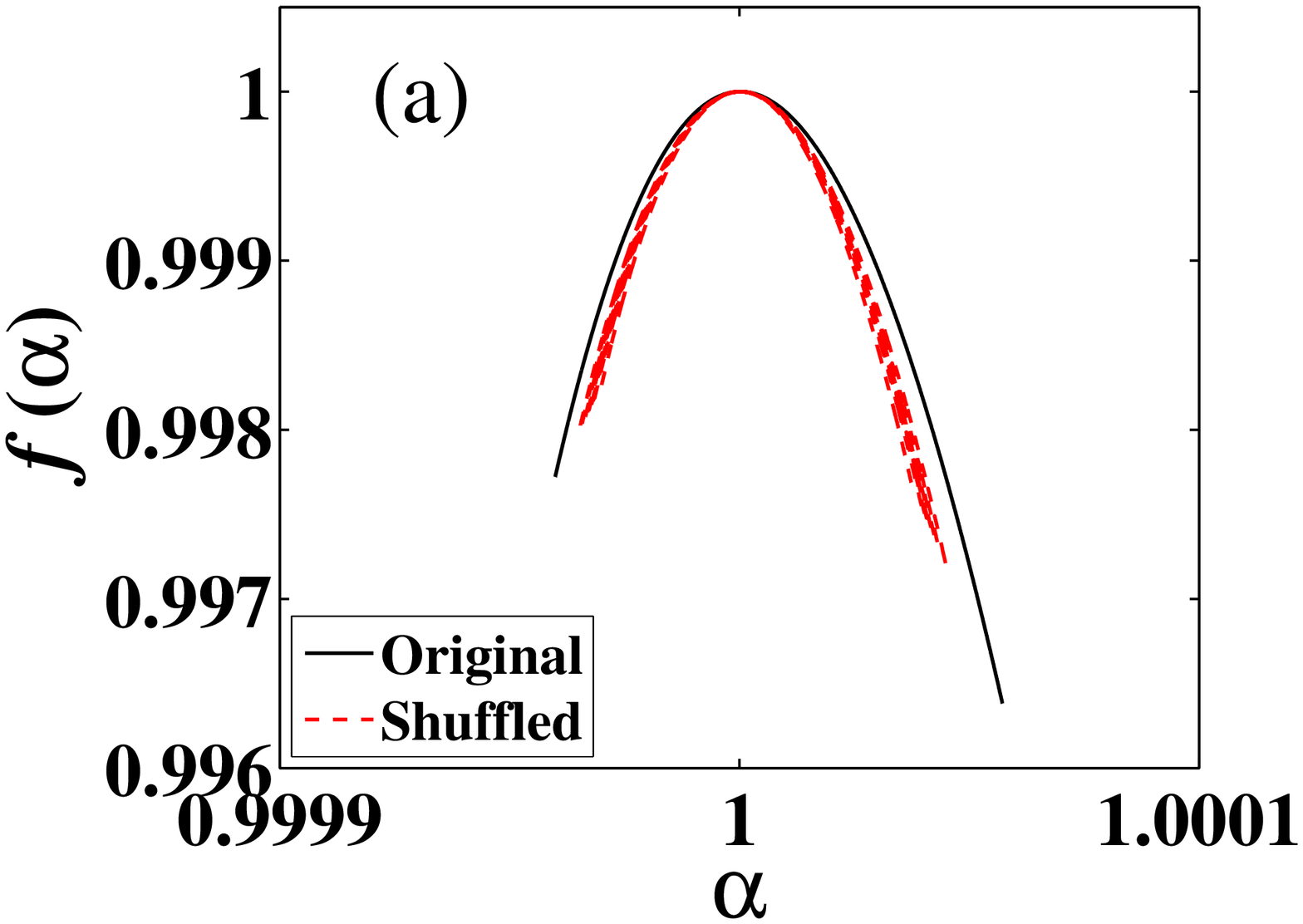}
\includegraphics[width=6cm]{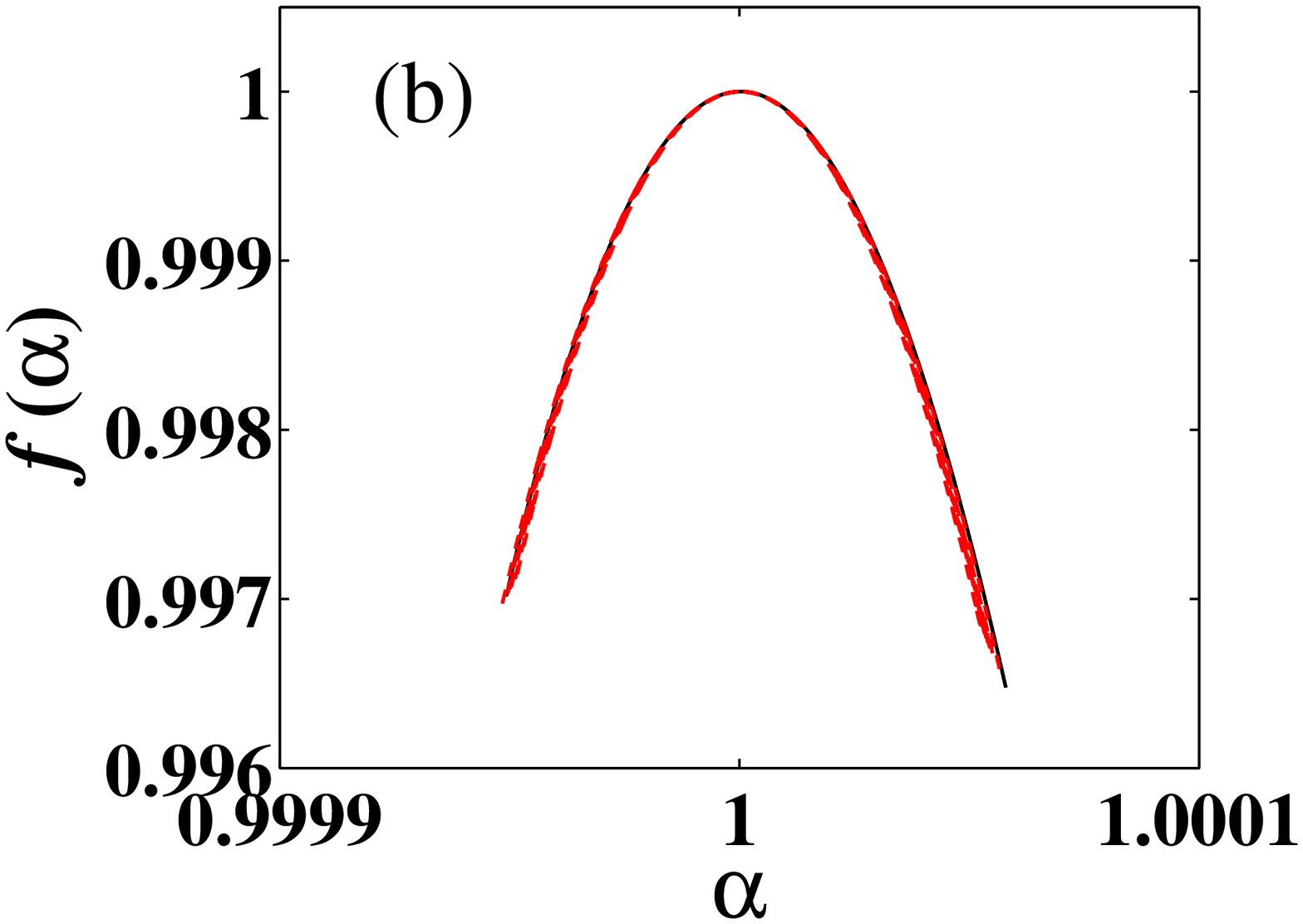}
\includegraphics[width=6cm]{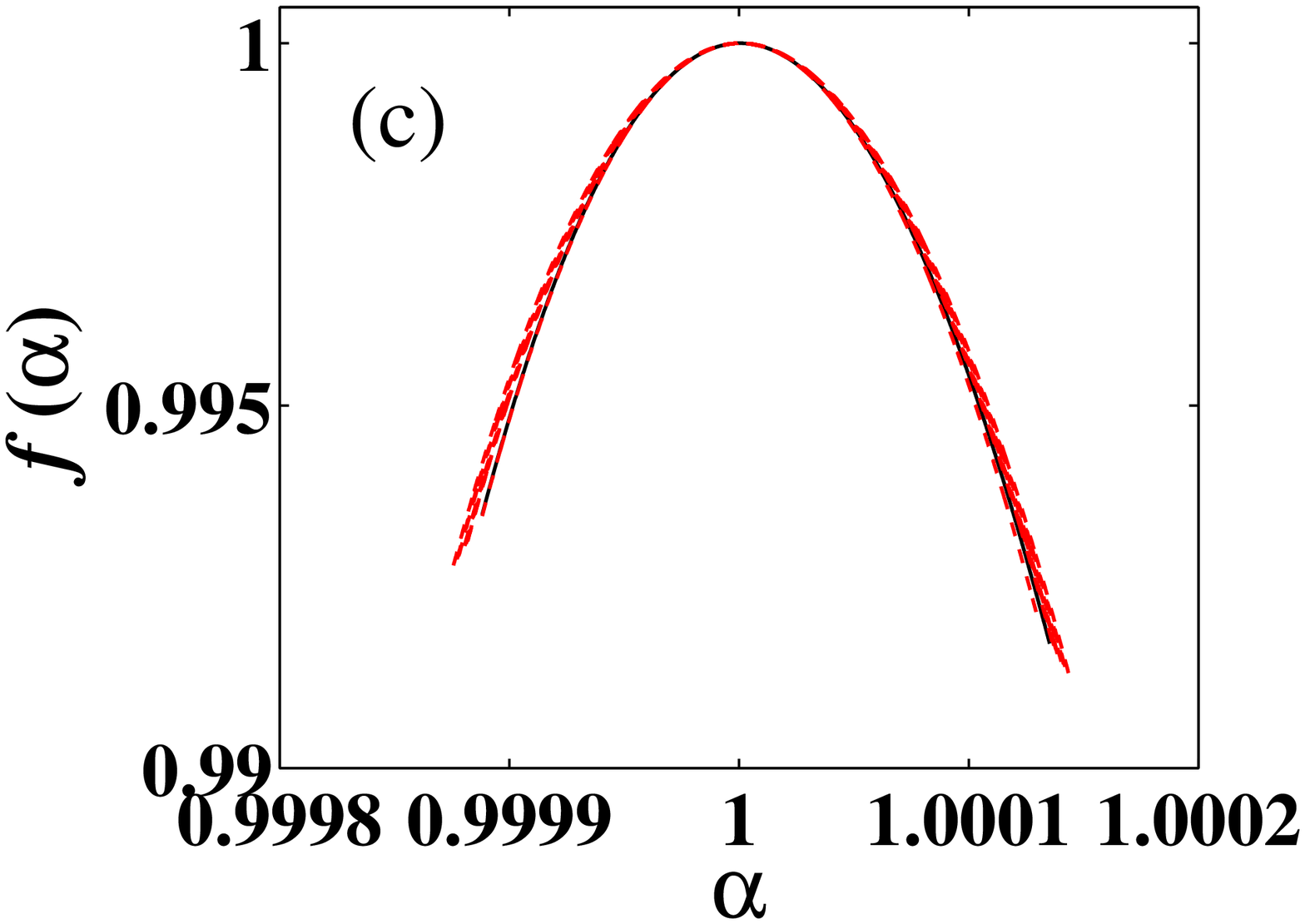}
\includegraphics[width=6cm]{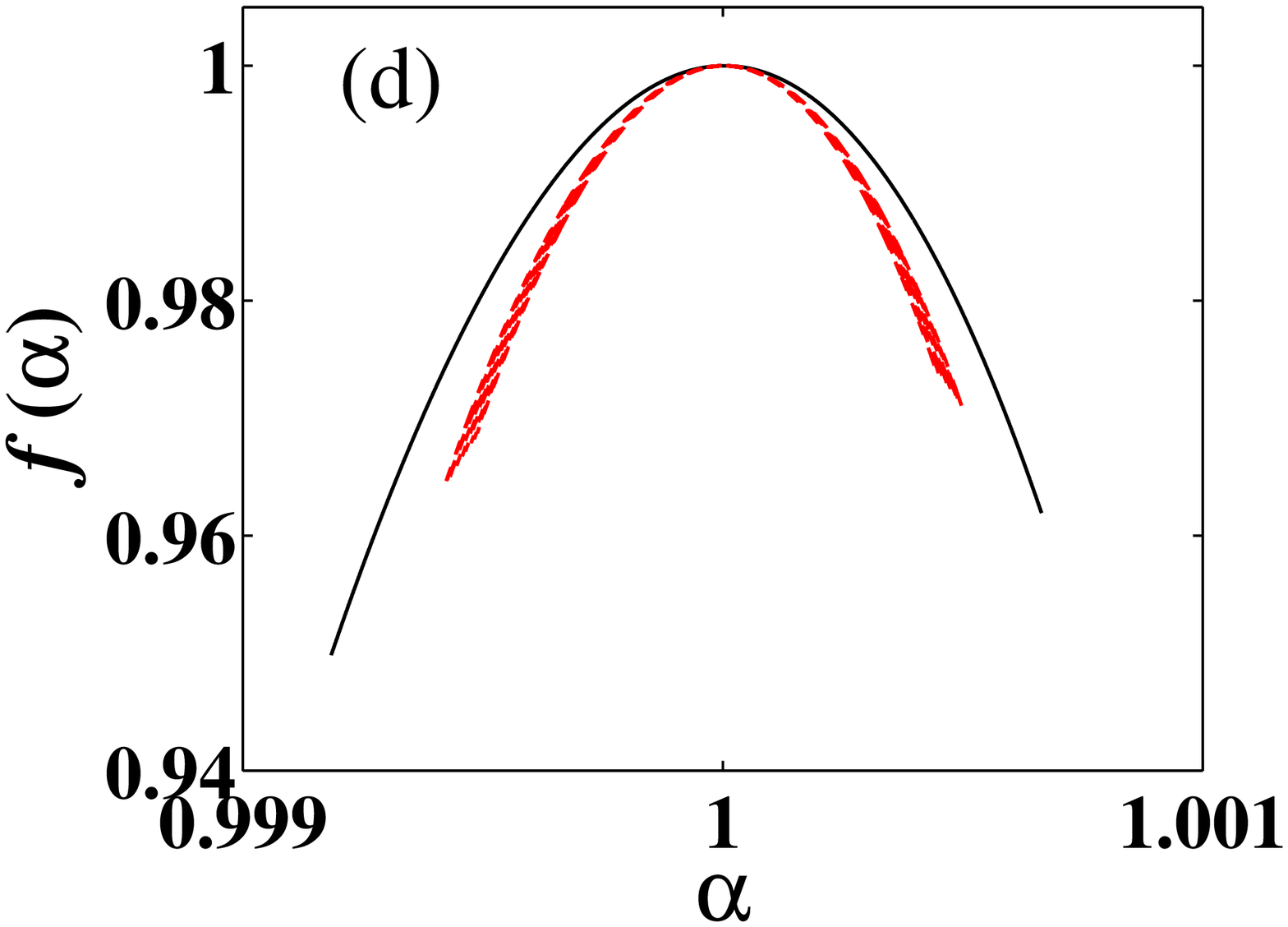}
\caption{(color online) Comparison of multifractal spectra extracted
from real and shuffled stock marker indexes. The solid lines are the
real data, while the dotted lines are the shuffled data. (a) HSI,
(b) SZSC, (c) S\&P 500, and (d) NASDAQ.} \label{Fig:SFalpha}
\end{figure}

We access further the statistical significance of the empirical
multifractality in the sprit of bootstrapping. For a given intraday
time series, we reshuffle the series to remove any potential
temporal correlation and carry out the same multifractal analysis as
for the original data. For the four examples discussed in
Sec.~\ref{s4:MA}, we compute the multifractal spectra of ten
reshuffled time series for each index. The results are illustrated
in Fig.~\ref{Fig:SFalpha}, where the solid lines are associated with
real stock market indexes, while the dotted lines are obtained from
the shuffled data of the corresponding indexes. We find that the
multifractal spectra of the real indexes $f(\alpha)$ and that of the
shuffled data $f_{\rm{rnd}}(\alpha_{\rm{rnd}})$ are almost
overlapping together in Fig.\ref{Fig:SFalpha} (b) and (c). Although
the solid lines and the dotted line can be distinguished clearly in
Fig.\ref{Fig:SFalpha} (a) and (d), the differences between $\alpha$
and $\alpha_{\rm{rnd}}$ are ignorable. In other words, the
multifractal nature in the real indexes is insignificant in these
examples.

\begin{figure}[htb]
\centering
\includegraphics[width=6cm]{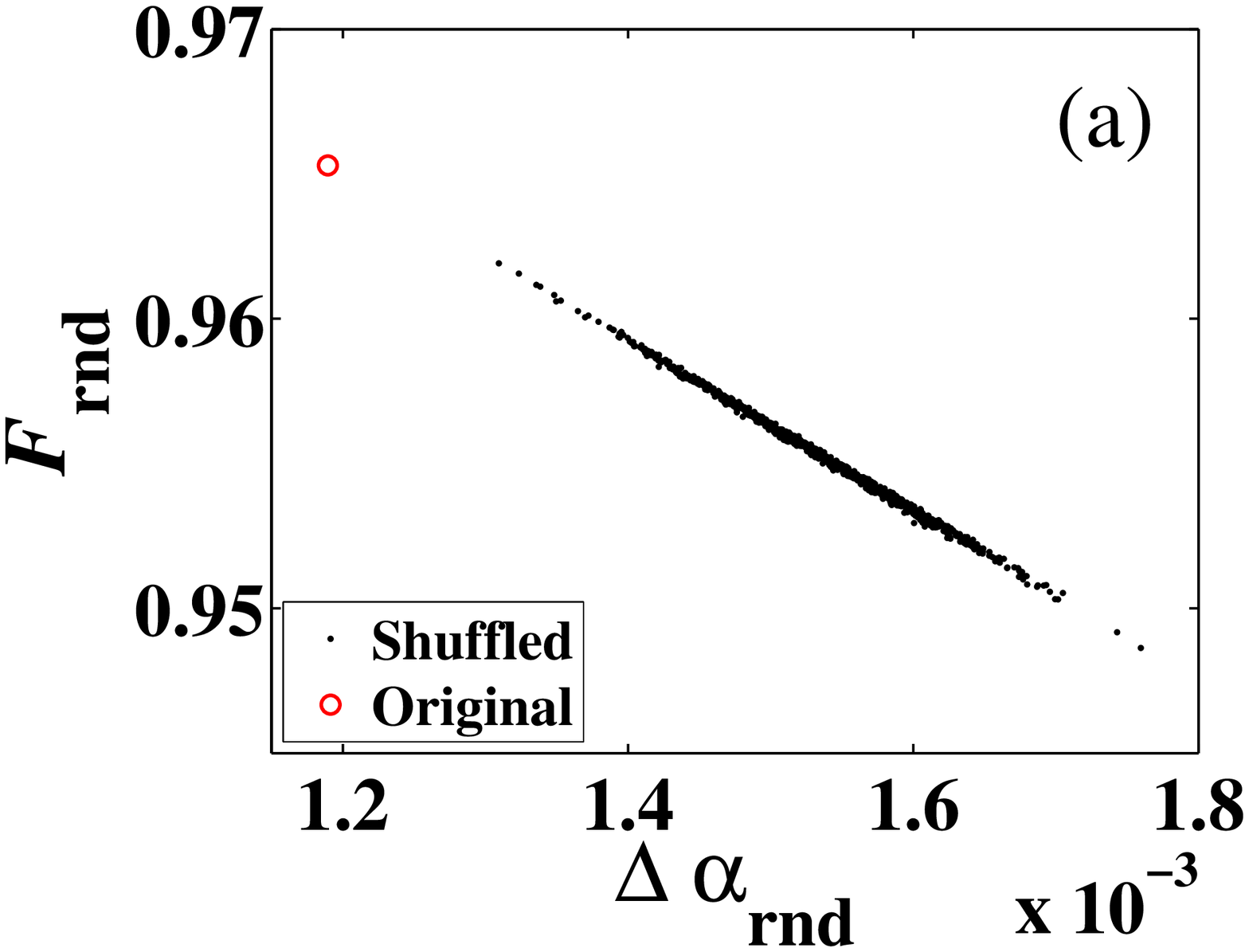}
\includegraphics[width=6cm]{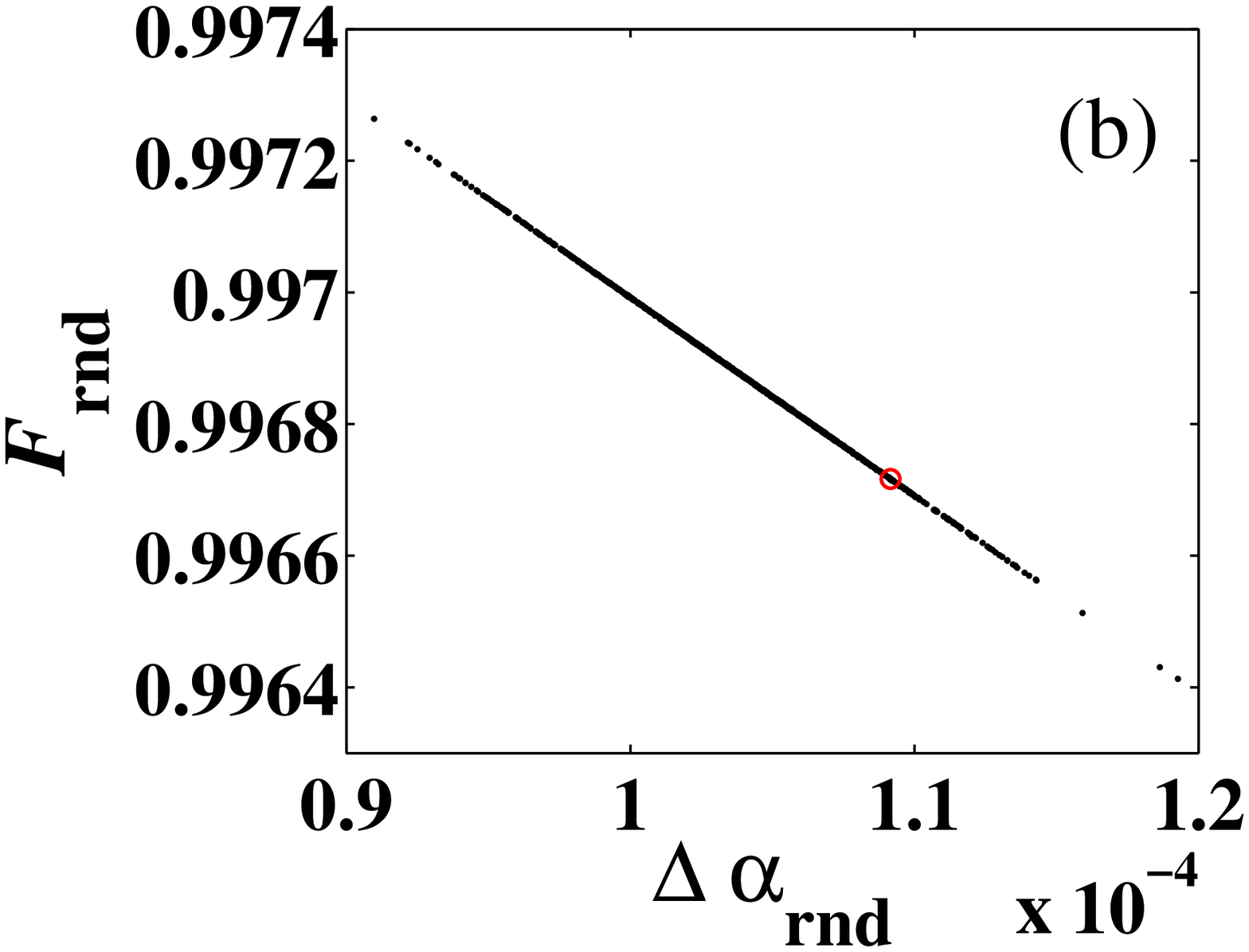}
\includegraphics[width=6cm]{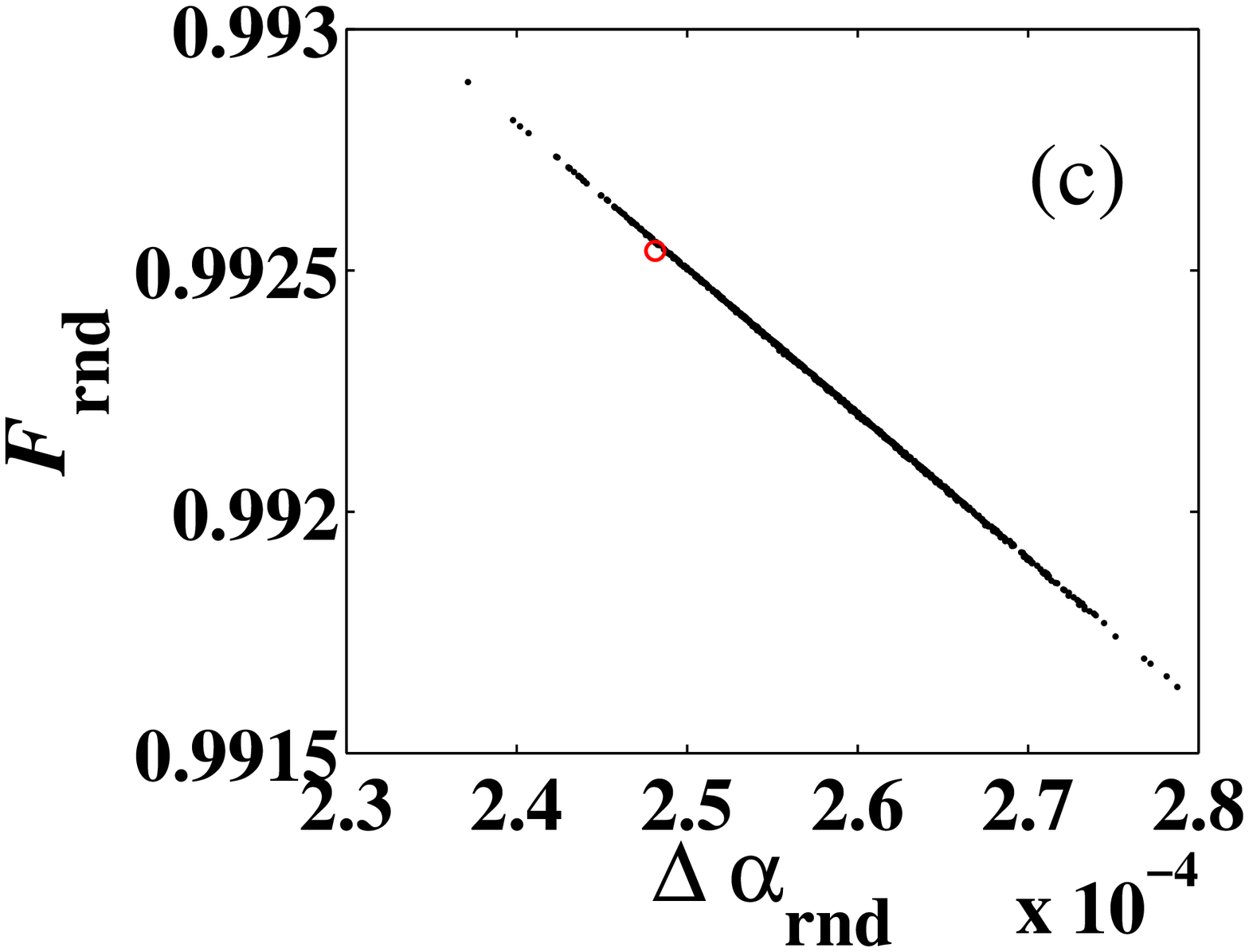}
\includegraphics[width=6cm]{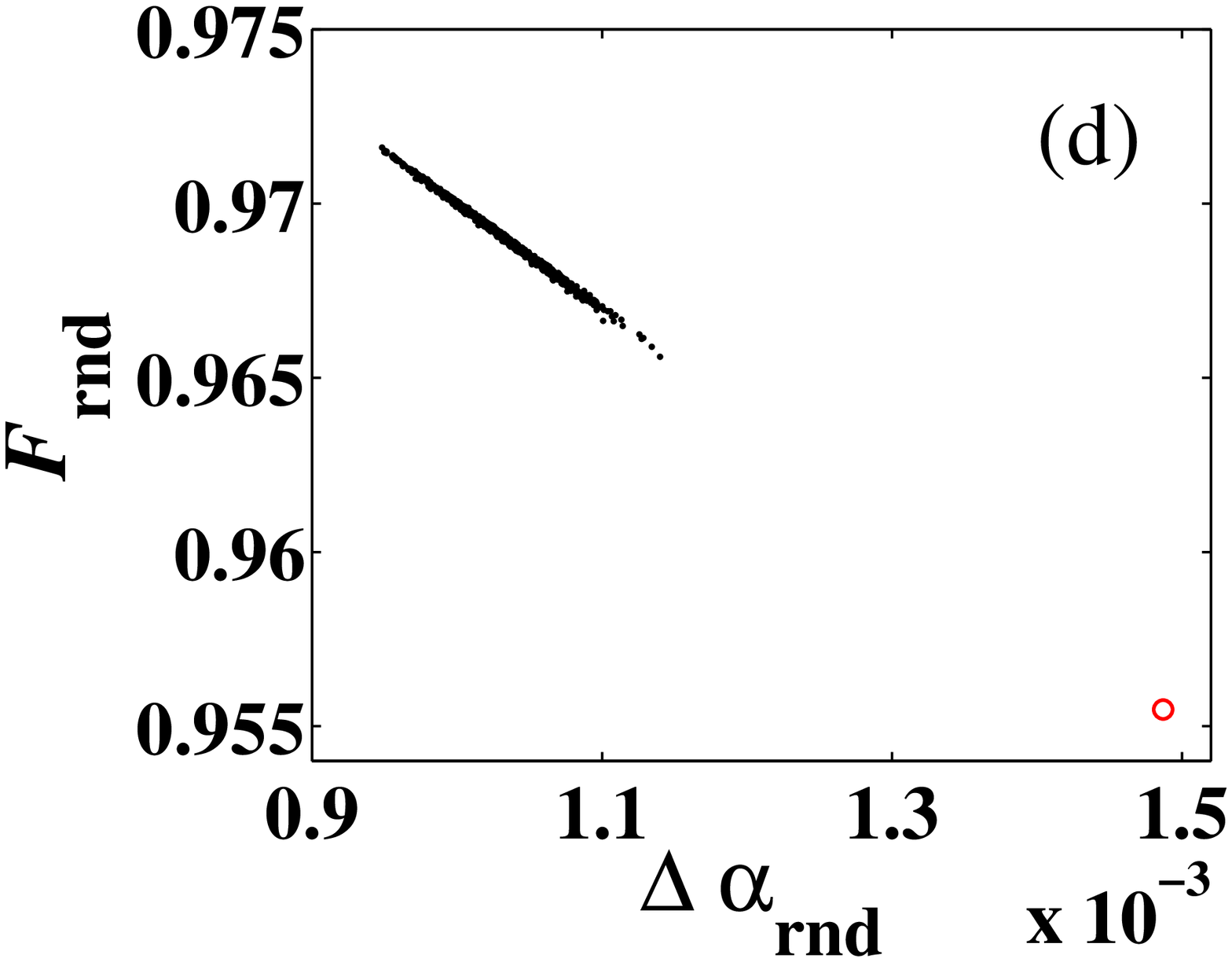}
\caption{Scatter plots of the dependence of the shuffled
$F_{\rm{rnd}}$ and the corresponding $\Delta \alpha_{\rm{rnd}}$. (a)
HSI, (b) SZSC, (c) S\&P 500, and (d) NASDAQ.} \label{Fig:scatter}
\end{figure}

For each intraday time series, we shuffle the data for 1000 times.
The associated multifractal spectra are obtained. For each
singularity spectrum, we calculate two characteristic quantity,
$\Delta\alpha$ and $F \triangleq [ f(\alpha_{\min}) +
f(\alpha_{\max}) ] / 2$. Figure~\ref{Fig:scatter} shows the scatter
plots of $F_{\rm{rnd}}$ for the shuffled data versus the
corresponding $\Delta \alpha_{\rm{rnd}}$ for the four example
trading days. Clear linear relationship between $F_{\rm{rnd}}$ and
$\Delta \alpha_{\rm{rnd}}$ for each case is observed and we have
\begin{equation}
F_{\rm{rnd}} = k \Delta \alpha_{\rm{rnd}} + b~,
 \label{Eq:linearity}
\end{equation}
where $k = -30.31$ and $b = 1.00$ for HSI, $k = -30.10$ and $b =
1.00$ for SZSC, $k = -30.05$ and $b = 1.05$ for S\&P 500, and $k =
-30.31$ and $b = 1.06$ for NASDAQ, respectively. The open circle in
each plot of Fig.~\ref{Fig:scatter} presents the values of $F$ and
$\Delta \alpha$ for the real data.

Two striking facts emerge from Fig.~\ref{Fig:scatter}. First, the
4000 points of $(\Delta\alpha_{\rm{rnd}}, F_{\rm{rnd}})$ collapse on
a same linear line since the values of $k$ and $b$ are identical for
the four plots. Second, the four points of $(\Delta\alpha, F)$ for
the four real data sets also locate on the same line. For other
trading days, we have observed similar phenomena, which put further
evidence on our conclusion that the real and reshuffled time series
have undistinguishable scaling behaviors.

\begin{figure}[htb]
\centering
\includegraphics[width=6cm]{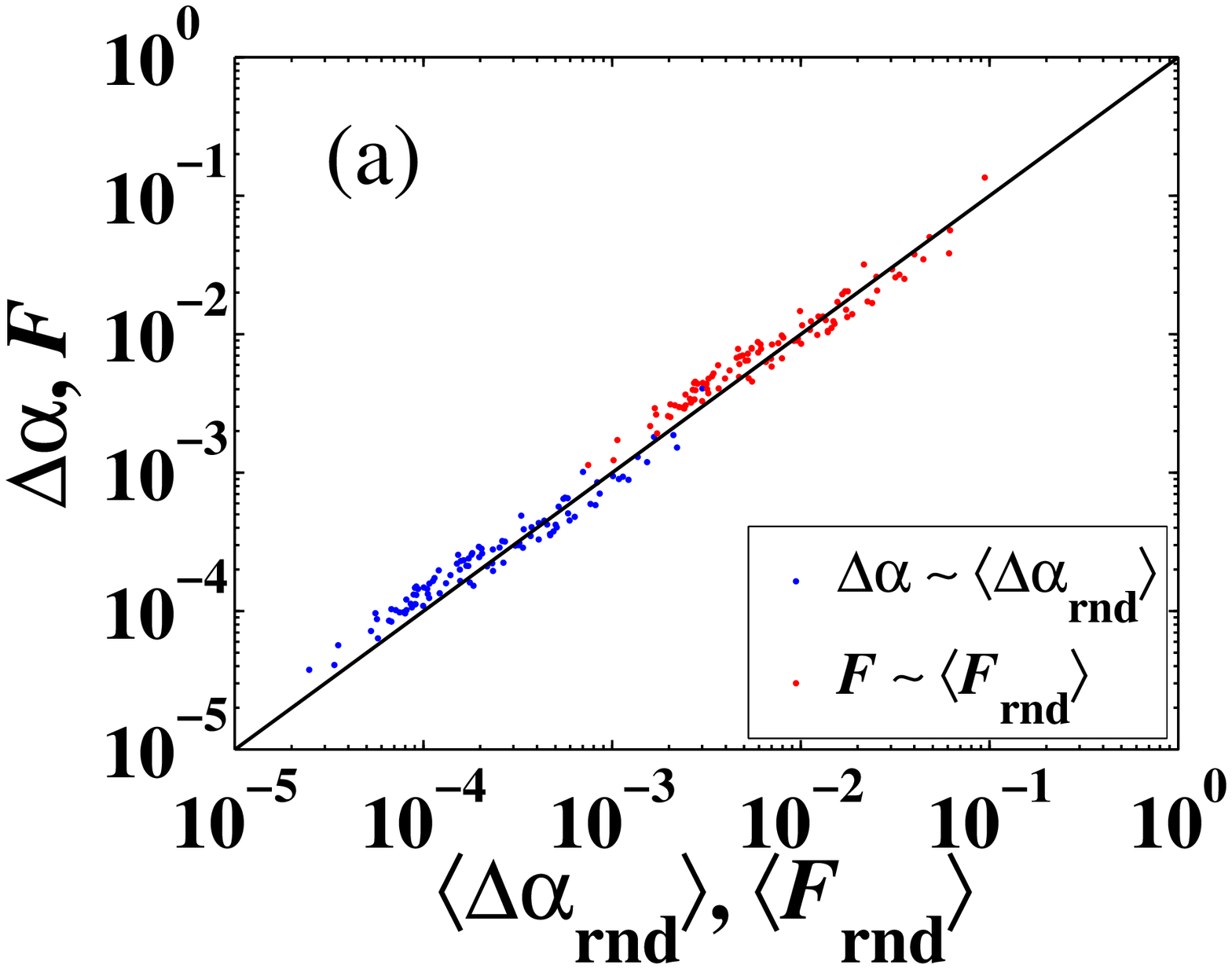}
\includegraphics[width=6cm]{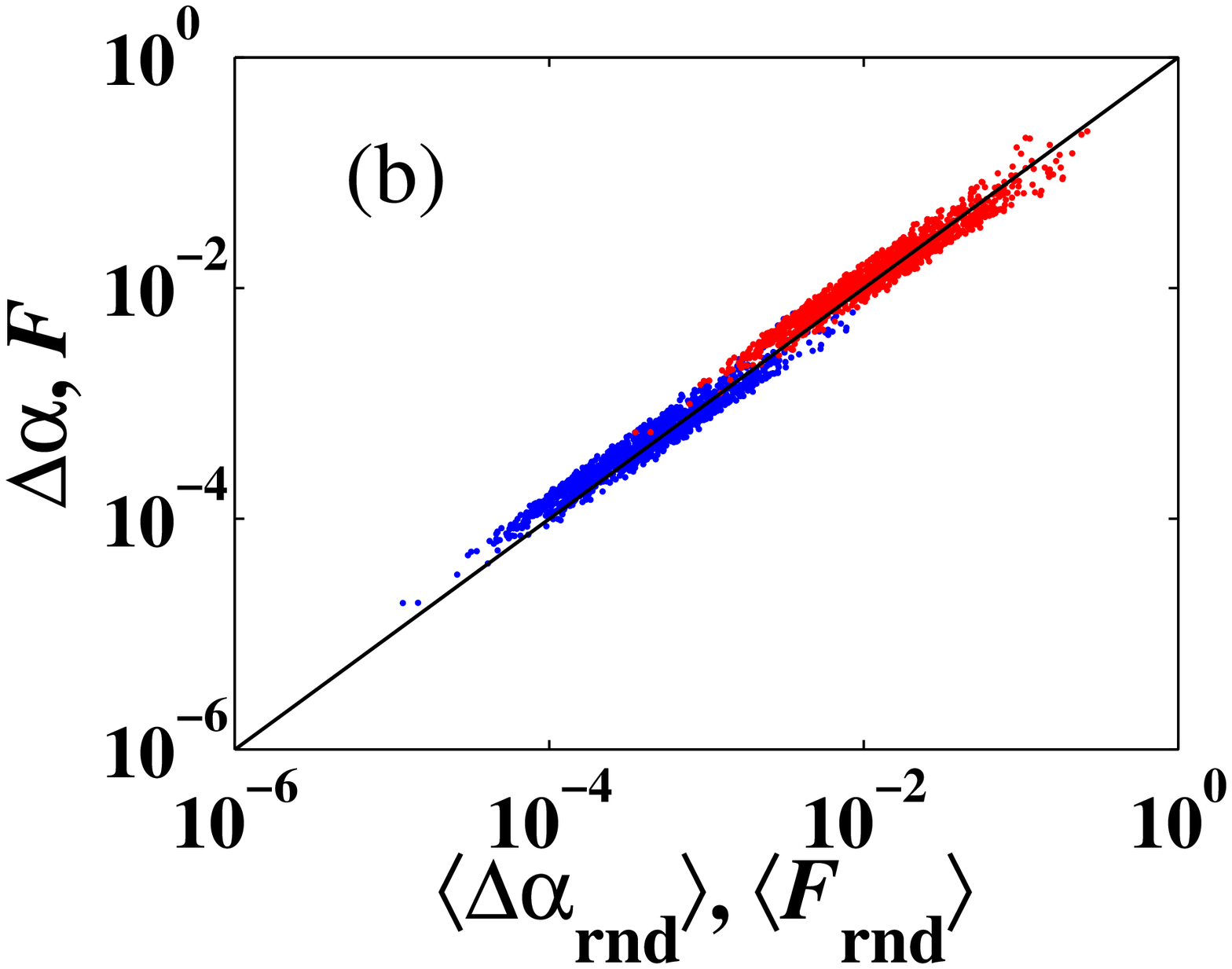}
\includegraphics[width=6cm]{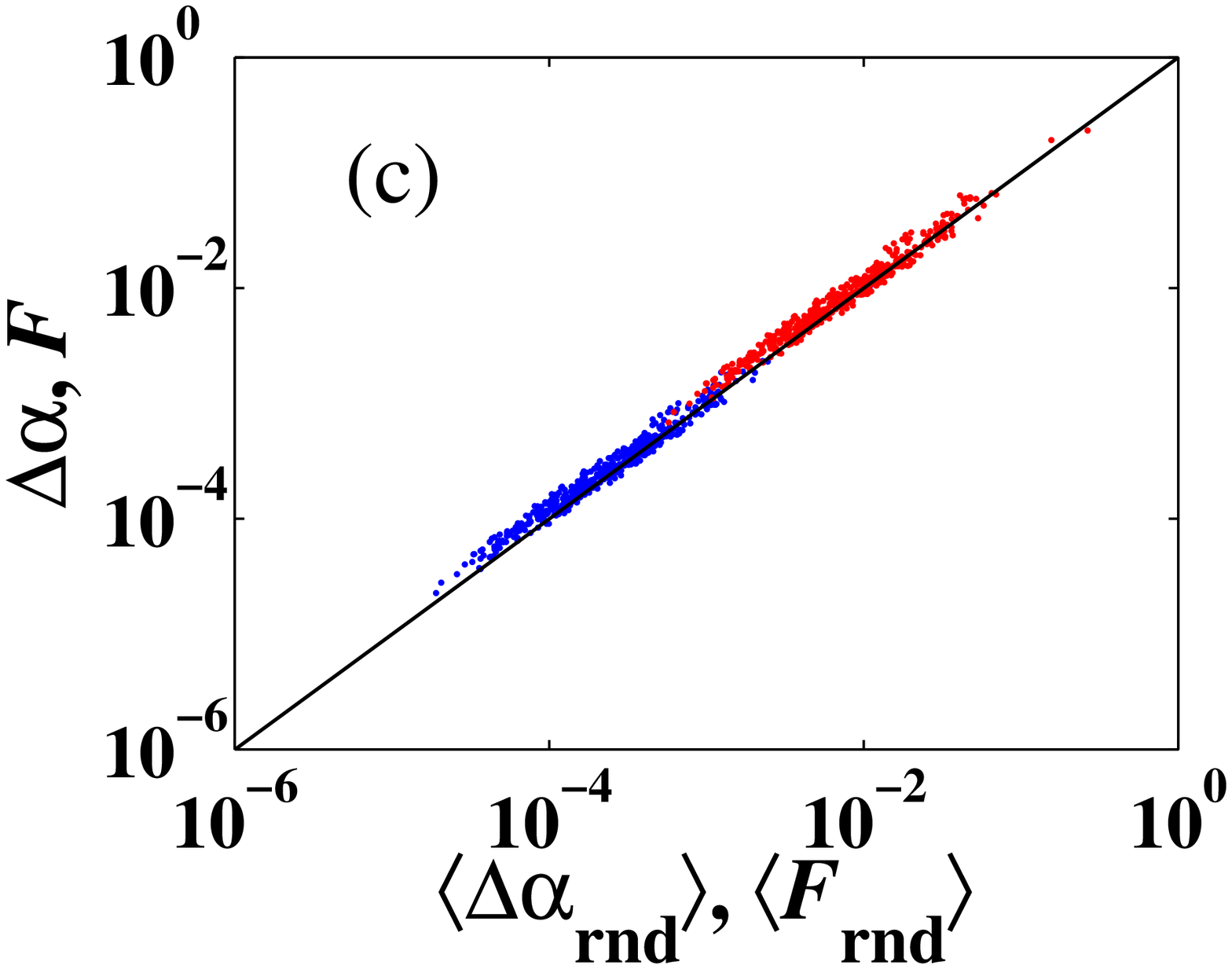}
\includegraphics[width=6cm]{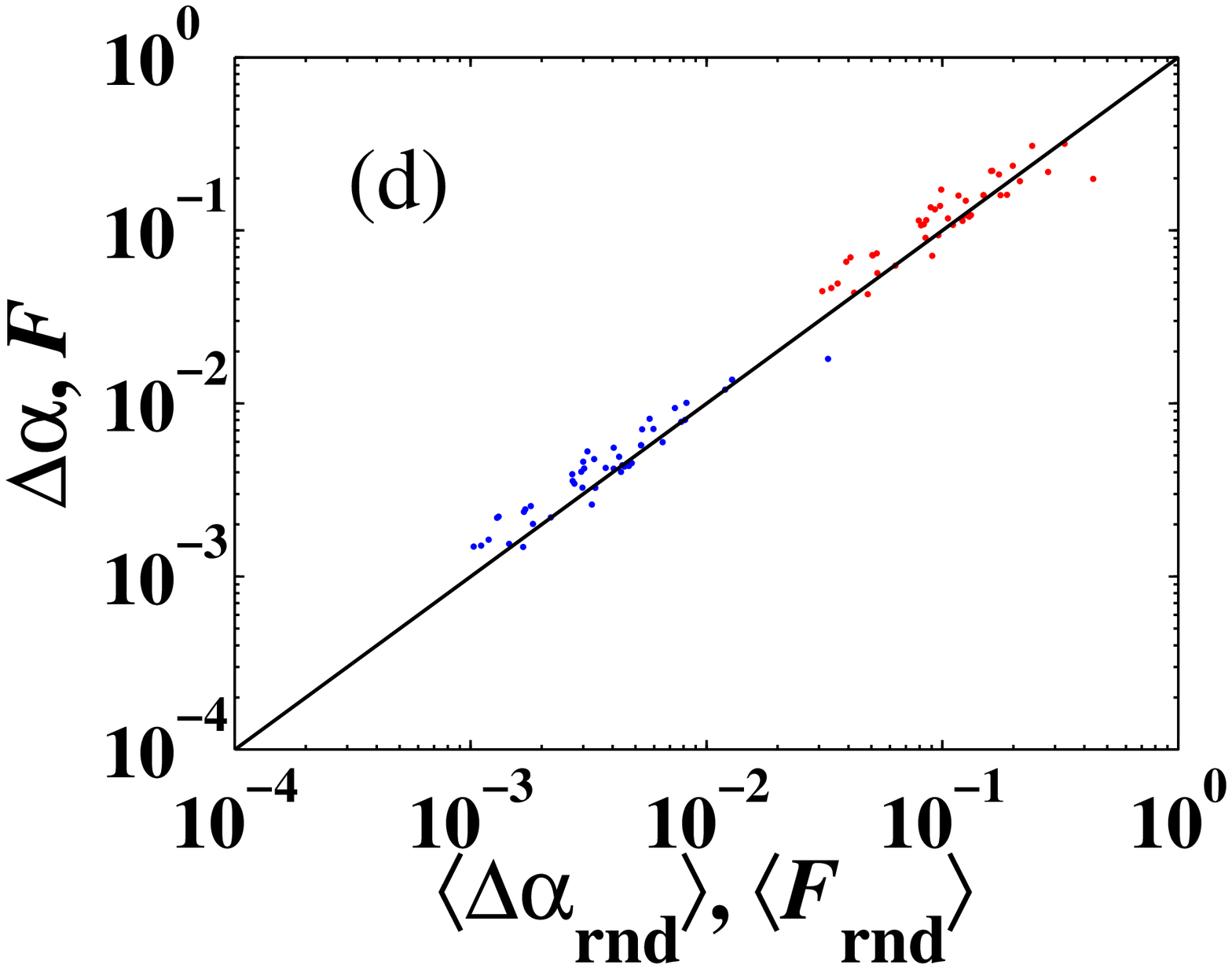}
\caption{Comparison of $F$ and $\Delta \alpha$ obtained from the
shuffled data and the real data. (a) HSI, (b) SZSC, (c) S\&P 500,
and (d) NASDAQ.} \label{Fig:RSC}
\end{figure}

The values of $\Delta\alpha$ and $F$ for each original time series
are compared with the averages $\langle \Delta \alpha_{\rm{rnd}}
\rangle$ and $\langle F_{\rm{rnd}} \rangle$ of the 1000
corresponding shuffled data sets. The results for the four indexes
are illustrated in Fig.~\ref{Fig:RSC}. The solid line is the main
diagonal $y = x$. We find that $\Delta \alpha \approx \langle \Delta
\alpha_{\rm{rnd}} \rangle$ and $F \approx \langle F_{\rm{rnd}}
\rangle$ for all cases, which implied that the multifractal spectra
of the shuffled data are very close to that of the real data and the
$f(\alpha)$ curves of real index data can be completely interpreted
by the random fluctuations of the original data sets. We stress that
there are no extreme values in the intraday index prices so that one
can not attribute the observed multifractality to tail fatness that
is absent in the present case. Hence, the multifractal property in
high-frequency stock market indexes obtained by partition function
method is not statistically significant. It is just an illusion.

In the presence case, to test the presence of multifractality
amounts to testing whether the local singularity exponent $\alpha
\neq 1$, or $\Delta \alpha \neq 0$. As a last step, we impose a very
strict null hypothesis to investigate whether the $f(\alpha)$
spectrum is wider than those produced by chance. The null hypothesis
is the following:
\begin{equation}
~~~~H_0: \Delta \alpha \leqslant \Delta \alpha_{\rm{rnd}}~.
\end{equation}
We can compute the $p$-value, which is the probability that the null
hypothesis is true. The smaller the $p$-value, the stronger the
evidence against the null hypothesis and favors the alternative
hypothesis that the presence of of multifractality is statistically
significant. The false probability is estimated by
\begin{equation}
p_1 = \Pr(\Delta \alpha \leqslant \Delta \alpha_{\rm{rnd}})~.
\end{equation}
Under the conventional significance level of $0.05$, the
multifractal phenomenon is statistically significant if and only if
$p_1 \leqslant 0.05$. While $p_1 > 0.05$, the null hypothesis cannot
be rejected. A similar null hypothesis can be described as follows:
\begin{equation}
H_0: F \geqslant F_{\rm{rnd}}~,
\end{equation}
where the false probability is
\begin{equation}
p_2 = \Pr(F \geqslant F_{\rm{rnd}})~.
\end{equation}
Using the conventional significance level of $0.05$, the
multifractal phenomenon is statistically significant if and only if
$p_2 \leqslant 0.05$.

For the four examples shown in Fig.~\ref{Fig:scatter}, we find that
$p_1 = 1$ and $p_2 = 1$ for HSI, $p_1 = 0.108$ and $p_2 = 0.109$ for
SZSC, $p_1 = 0.452$ and $p_2 = 0.456$ for S\&P 500, and $p_1 = 0$
and $p_2 = 0$ for NASDAQ. Obviously, we can not distinguish the real
data from the shuffled data beside NASDAQ for the chosen trading
days. We also find that $p_1 \approx p_2$ for all the trading days.
More generally, Table~\ref{Tb:1} shows the statistical tests for the
all the each trading days. About half of the trading days can not
pass the statistical inference, indicating that multifractality is
absent in the those trading series.

\begin{table}[htp]
\begin{center}
\caption{\label{Tb:1} Statistical tests for the presence of
multifractal nature in the four indexes investigated.}
\medskip
\begin{tabular}{ccccc}
  \hline\hline
   Indexes  & HSI & SZSC & S\&P 500 & NASDAQ  \\\hline
  Percentage of $p_1 \leqslant 0.05$ & 54.6\% & 56.1\% & 54.4\% & 53.9\%  \\
  Percentage of $p_2 \leqslant 0.05$ & 54.4\% & 55.8\% & 53.6\% & 53.6\%  \\
  \hline\hline
\end{tabular}
\end{center}
\end{table}

\section{Conclusion}
\label{s6:con}

We have investigated the multifractal features in intraday minutely
high-frequency stock market indexes (including HSI, SZSC, S\&P 500,
and NASDAQ) for individual trading days. The resultant scaling
functions $\tau(q)$ have been confirmed to be linear and the
singularities $\alpha$ are close to 1 so that $\Delta\alpha$ is
close to 0. This analysis implies that there is no multifractality
in the indexes. Further evidence based on bootstrapping technique
shows that that the scaling behavior of the shuffled data is
undistinguishable from that of the raw data. Specifically, we find
that, (1) almost all points $(\Delta\alpha,F)$ of the raw data sets
locate on the same straight line $F_{\rm{rnd}} = -30 \Delta
\alpha_{\rm{rnd}} + 1$ extracted from the points $(\Delta
\alpha_{\rm{rnd}},F_{\rm{rnd}})$ of the shuffled data; (2) for each
time series, $\Delta \alpha \approx \langle \Delta \alpha_{\rm{rnd}}
\rangle$ and $F \approx \langle F_{\rm{rnd}} \rangle$; and (3) the
two rather strict null hypotheses cannot be rejected for about half
of the time series. There is thus no doubt that the reported
multifractal nature in the indexes of HSI and SZSC
\cite{Sun-Chen-Wu-Yuan-2001-PA,Sun-Chen-Yuan-Wu-2001-PA,Wei-Huang-2005-PA}
is not a fact but a fiction. This conclusion is further verified by
two indexes (S\&P 500 and NASDAQ) in a developed stock market. We
believe that our analysis and conclusion apply for other market
indexes or common stock prices when one concerns intraday stock
prices or indexes rather than their returns.

In addition, we cast doubts on the efforts to use this illusionary
multifractal feature to forecast the stock market
\cite{Sun-Chen-Yuan-Wu-2001-PA} and to define a risk index for risk
management \cite{Wei-Huang-2005-PA}. However, to be more
conservative, we do not deny the potential usefulness of those
techniques proposed based on some nonexistent properties. The idea
to use multifractal nature to predict or to manage risks in stock
markets should be investigated based on the returns or other
alternative financial quantities. After all, one cannot build a
palace on a sand beach.

\bigskip
{\textbf{Acknowledgments:}}

We are indebted to Prof. Bing-Hong Wang for providing the HSI data
and fruitful discussion. This work was partly supported by the
National Natural Science Foundation of China (Grant No. 70501011),
the Fok Ying Tong Education Foundation (Grant No. 101086),  and the
Shanghai Rising-Star Program (No. 06QA14015).

\bibliography{E:/Papers/Auxiliary/Bibliography}

\end{document}